\documentclass{article}

\usepackage{arxiv}
\usepackage[utf8]{inputenc} 
\usepackage[T1]{fontenc}    
\usepackage{hyperref}       
\usepackage{url}            
\usepackage{booktabs}       
\usepackage{amsfonts}       
\usepackage{nicefrac}       
\usepackage{microtype}      
\usepackage{lipsum}
\usepackage{graphicx}
\usepackage{float}
\usepackage[labelfont=bf]{caption}

\title{A flat persistence diagram for improved visualization of persistent homology}

\author{
  Raoul R. Wadhwa \\
  Cleveland Clinic Lerner College of Medicine \\
  Case Western Reserve University \\
  Cleveland, OH 44195, United States
  \And
  Andrew Dhawan \\
  Neurological Institute \\
  Cleveland Clinic \\
  Cleveland, OH 44195, United States
  \And
  Drew F.K. Williamson \\
  Department of Pathology \\
  Brigham and Women's Hospital \\
  Boston, MA 02115, United States
  \And
  Jacob G. Scott \\
  Translational Hematology and Oncology Research \\
  Cleveland Clinic \\
  Cleveland, OH 44195, United States \\
  \url{https://www.lerner.ccf.org/thor/scott/lab/}
}

\begin{document}

\maketitle

\begin{abstract}
Visualization in the emerging field of topological data analysis has progressed from persistence barcodes and persistence diagrams to display of two-parameter persistent homology.
Although persistence barcodes and diagrams have permitted insight into the geometry underlying complex datasets, visualization of even single-parameter persistent homology has significant room for improvement.
Here, we propose a modification to the conventional persistence diagram - the flat persistence diagram - that more efficiently displays information relevant to persistent homology and simultaneously corrects for visual bias present in the former.
Flat persistence diagrams display equivalent information as their predecessor, while providing researchers with an intuitive horizontal reference axis in contrast to the usual diagonal reference line.
Reducing visual bias through the use of appropriate graphical displays not only provides more accurate, but also deeper insights into the topology that underlies complex datasets.
Introducing flat persistence diagrams into widespread use would bring researchers one step closer to practical application of topological data analysis.
\end{abstract}


\section{Introduction}
\label{sec:intro}

Topological data analysis aims to characterize the shape of high-dimensional data.
However, given human inability to perceive more than three dimensions, gaining an intuitive geometric comprehension of high-dimensional datasets is a challenging task.
Persistent homology extracts topological features from complex datasets without requiring dimension reduction, although techniques such as UMAP can be used to ease computation~\citep{umap}.
Researchers have devised visualizations to display persistent homology as calculated from mathematical structures, such as Vietoris-Rips complexes, to provide a better understanding of complex datasets.
The most successful of these are the persistence barcode~\citep{barcode,barcode2} and the persistence diagram~\citep{persist-diag}.
Here, we propose a modification to improve conventional persistence diagrams for more effective visualization of persistent homology.

\section{Methods}
\label{sec:methods}

All persistent homology calculations and visualizations were completed using the \texttt{TDAstats} package \texttt{v0.4.0} for the \texttt{R} programming language \texttt{v3.5.1}~\citep{tdastats,r}.
All calculations and visualizations are fully reproducible, with relevant code available at \url{https://github.com/rrrlw/visual-tda}.
In this report, the cycle and feature nomenclatures are used interchangeably; thus, we use 0-cycle and dimension 0 feature interchangeably.

\section{Review of current visualizations}
\label{sec:review}

\begin{figure}
    \centering
    \begin{minipage}{0.49\textwidth}
        \centering
        \includegraphics[width=0.8\textwidth]{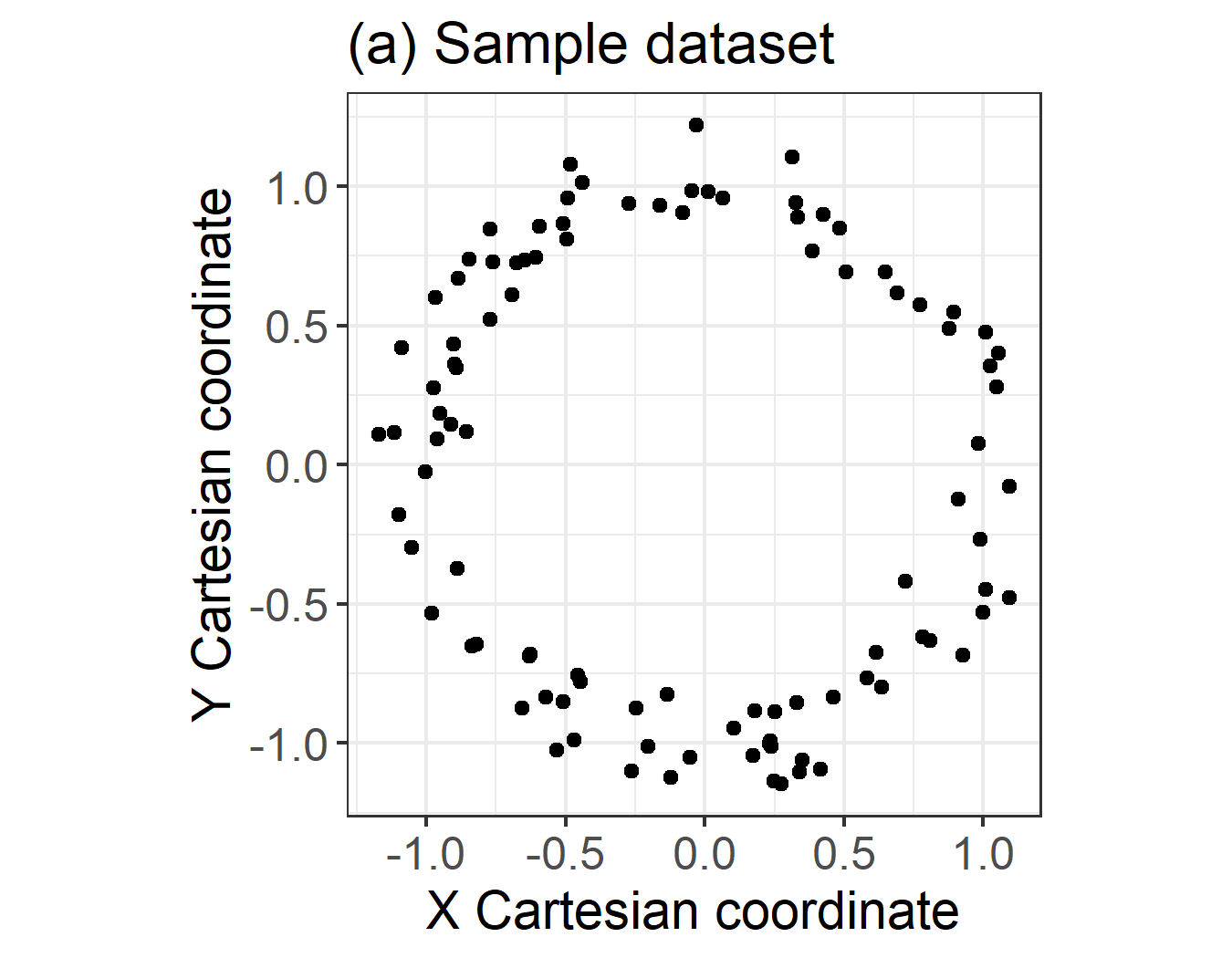}
    \end{minipage}
    \begin{minipage}{0.49\textwidth}
        \centering
        \includegraphics[width=0.8\textwidth]{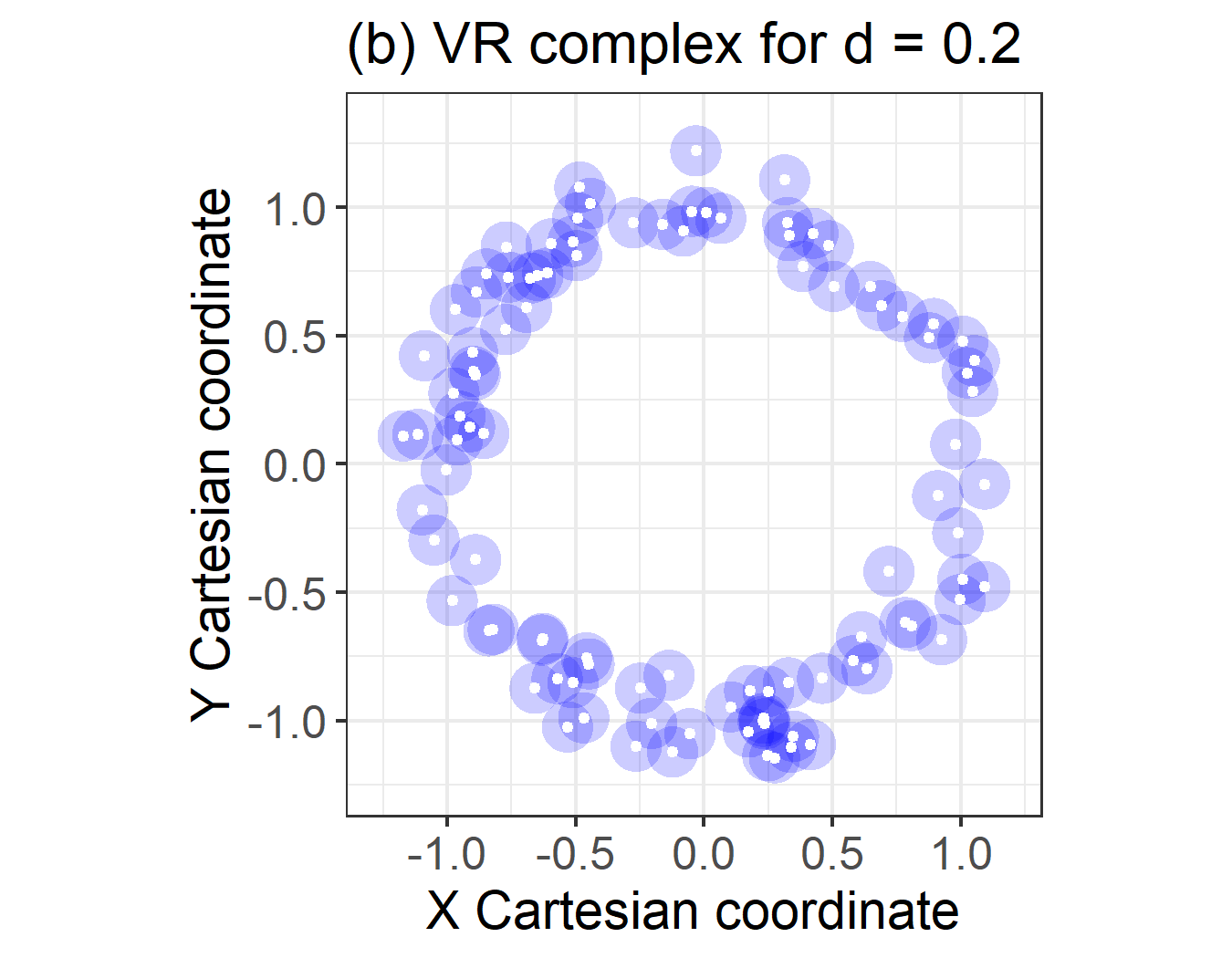}
    \end{minipage}
    
    \begin{minipage}{0.49\textwidth}
        \centering
        \includegraphics[width=0.8\textwidth]{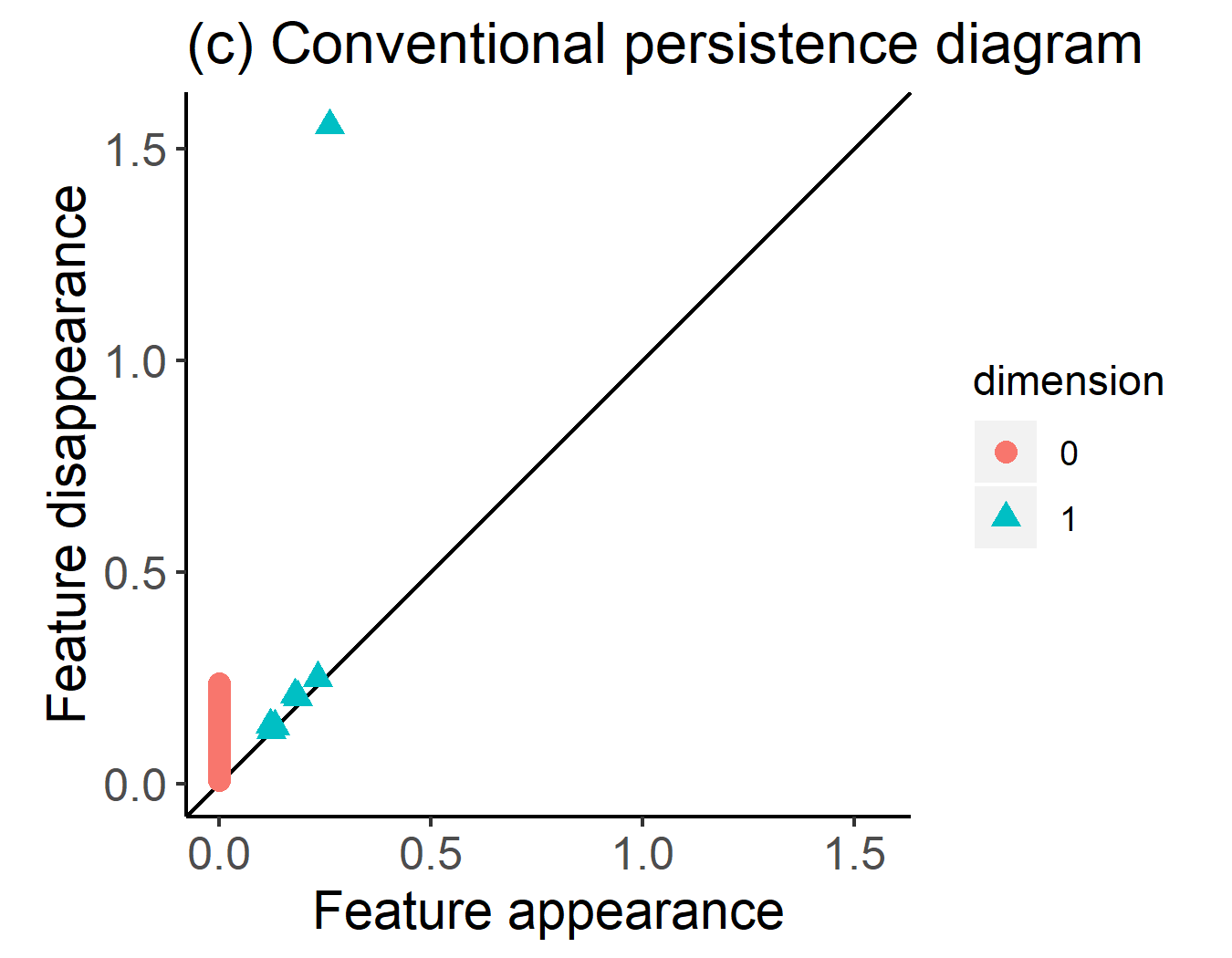}
    \end{minipage}
    \begin{minipage}{0.49\textwidth}
        \centering
        \includegraphics[width=0.8\textwidth]{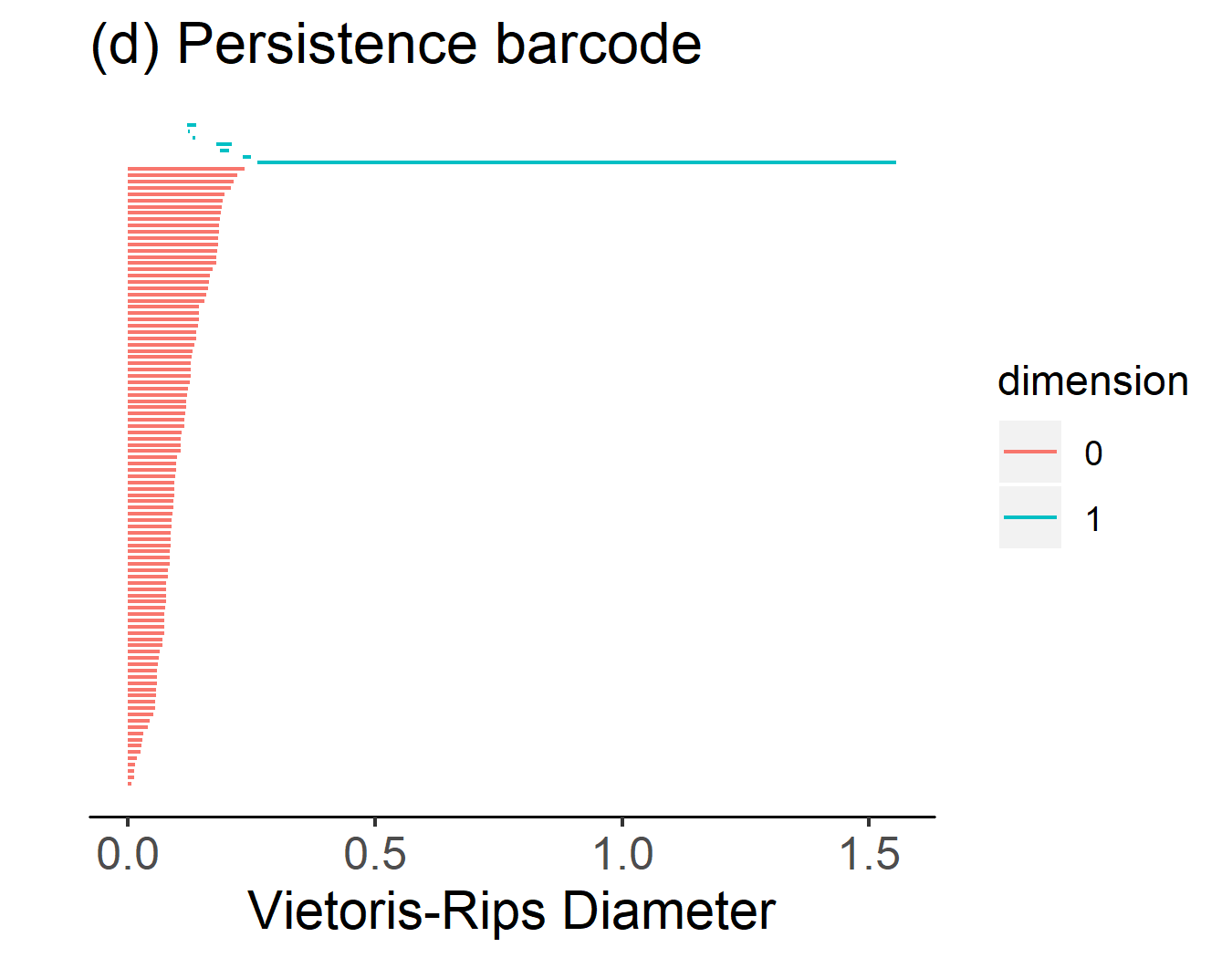}
    \end{minipage}
    \caption{\textbf{Visualizing persistent homology of an annulus.} Panel a: Scatterplot of 100 points uniformly selected from the circumference of a noisy unit circle. Panel b: Corresponding Vietoris-Rips complex with $d=0.2$. A circle with diameter equal to $d$ is drawn at each point in (a). Edges are drawn between the centers of overlapping circles, forming a simplicial complex for a single value of $d$. Panel c: Conventional persistence diagram for point cloud in (a). The single 1-dimensional feature (blue triangle) represents the dominant 1-cycle in an annulus. Panel d: Persistence barcode for point cloud in (a). The single 1-dimensional interval (blue) represents the dominant 1-cycle present in an annulus. The 1-cycle first appears at $d=0.26$, corresponding to the x-coordinate of the dimension 1 feature in (c) and the left boundary of the dimension 1 interval in (d); this is consistent with (b), where $d=0.2$ falls slightly short of connecting each point to a neighbor, which would complete the 1-cycle. Every point in (c) corresponds to exactly one interval in (d). However, visual confirmation of this fact is complicated by overlap of dimension 0 features in (c).}
    \label{fig:examples}
\end{figure}

A persistence diagram (Figure~\ref{fig:examples}c) is a set of points in the first quadrant of a 2-dimensional Cartesian space.
Each point corresponds to a single feature, where the first coordinate ($x$) equals the Vietoris-Rips diameter ($d$) at feature appearance and the second coordinate ($y$) equals $d$ at feature disappearance.
A reference line ($y=x$) is often drawn to permit visual calculation of feature persistence ($y-x$) as the vertical distance between a point and the reference line.
A persistence barcode (Figure~\ref{fig:examples}d) is a set of intervals on the non-negative subset of the 1-dimensional real number line.
Each interval corresponds to a single feature, where its left border equals $d$ at feature appearance and the right border equals $d$ at feature disappearance.
Thus, there is a bijection between points in the persistence diagram representation of persistent homology and intervals in the corresponding persistence barcode representation.

Since $x<y$ must hold true, the triangular half of the persistence diagram under the reference line is always unused whitespace, an undesired feature in effective graphics design~\citep{tufte2001}.
Since persistence diagrams also look most aesthetically appropriate when the horizontal and vertical axes share the same range, features with large $y$ can often create further unnecessary whitespace by increasing the maximum border of $x$.
This is clear in Figure~\ref{fig:examples}c where everything to the right of $x=0.26$ is unnecessary whitespace that serves primarily to satisfy the aesthetic requirement of a square persistence diagram.
Additionally, points in persistence diagrams naturally overlap for similar features making some aspects of the diagram inherently unclear (red circles in Figure~\ref{fig:examples}c).
Although the issue is avoided completely in persistence barcodes, where features are visually non-overlapping, conventional persistence diagrams would benefit from more efficient use of space that would at least marginally decrease the degree of feature overlap.
Additionally, the persistence barcode fails to provide sufficient visual discrimination of individual intervals when an abundance of features is present - the main advantage of representing features as smaller components, i.e. as points in persistence diagrams - which motivates the search for improvements to the conventional persistence diagram.

\section{Flat persistence diagram}
\label{sec:proposal}

\begin{figure}
    \centering
    \begin{minipage}{0.33\textwidth}
        \includegraphics[width=\textwidth]{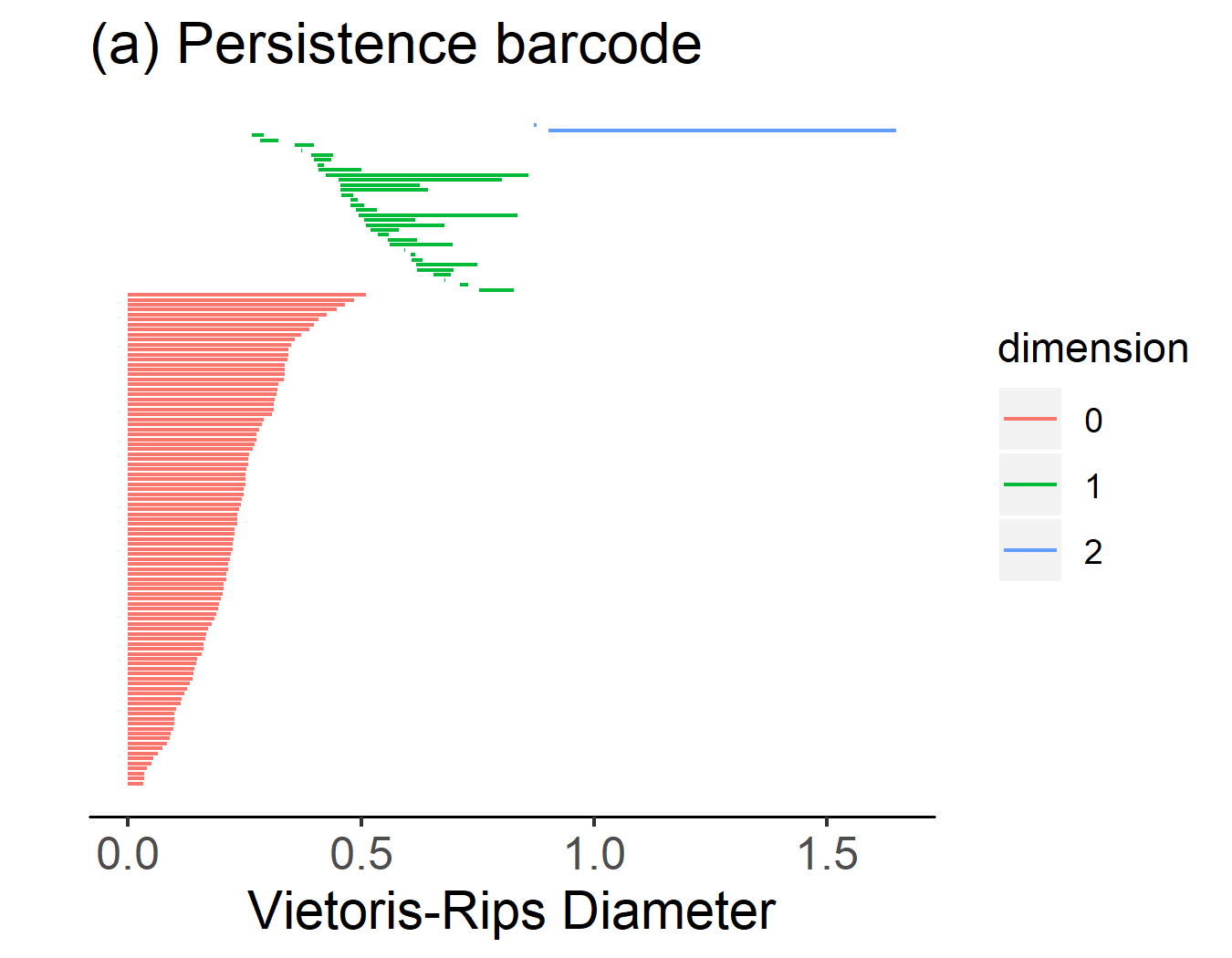}
    \end{minipage}
    \begin{minipage}{0.33\textwidth}
        \includegraphics[width=\textwidth]{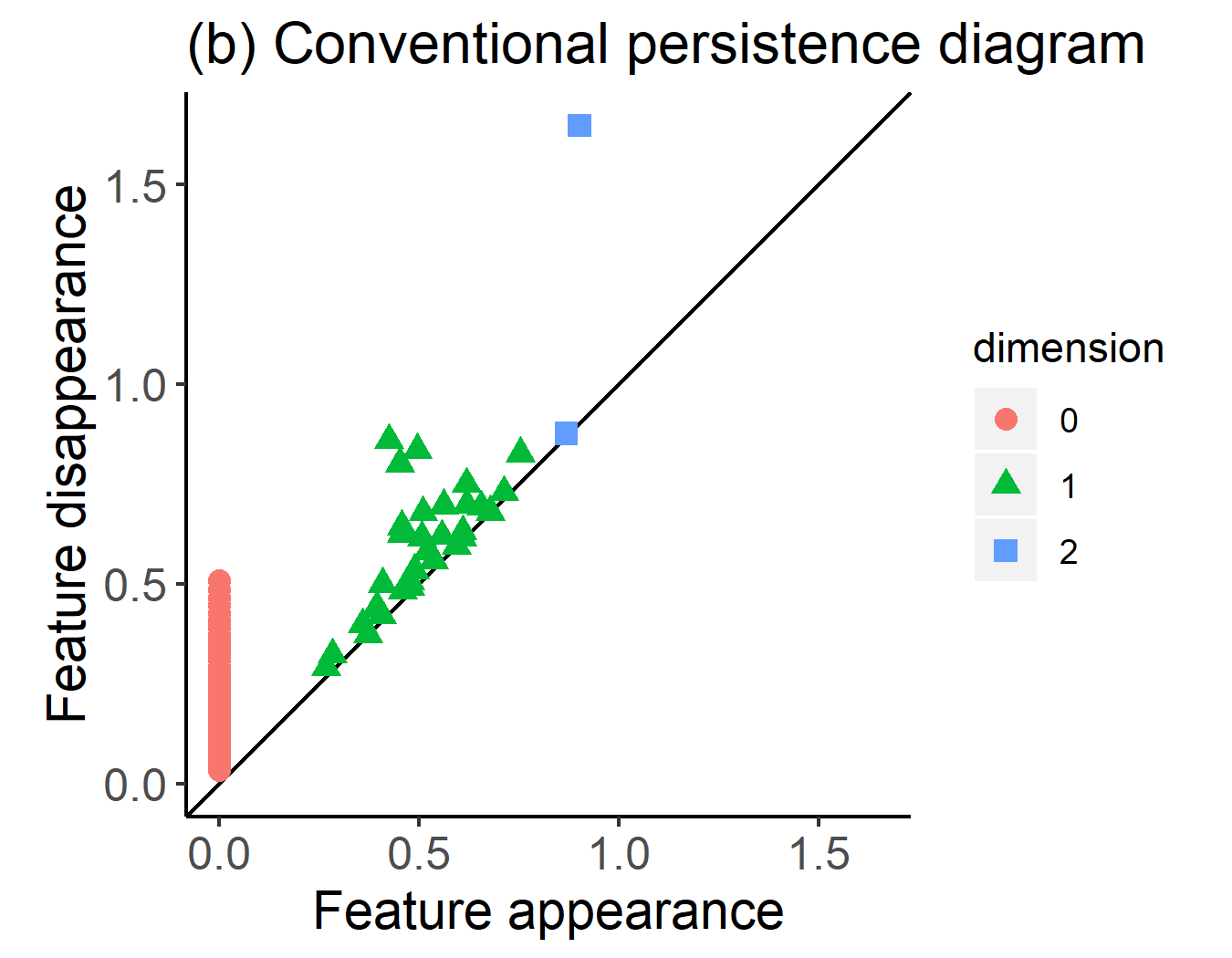}
    \end{minipage}
    \begin{minipage}{0.33\textwidth}
        \includegraphics[width=\textwidth]{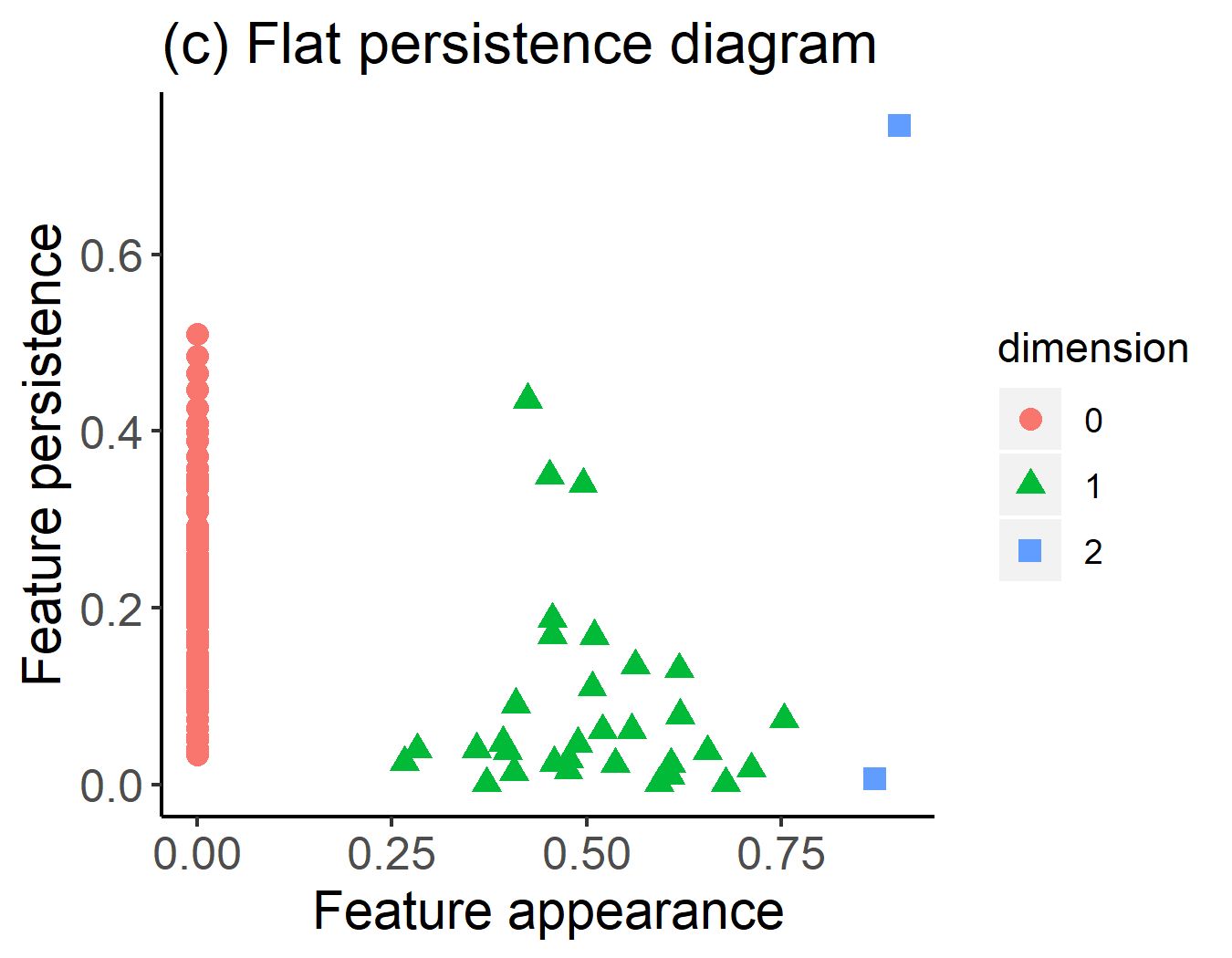}
    \end{minipage}
    \caption{\textbf{Persistent homology of the unit sphere.} Each panel plots persistent homology for the same 100 pseudorandom points selected from the surface of a unit sphere. Panel a: persistence barcode that clearly highlights the single dominant 2-cycle (long blue interval) expected for a hollow sphere. Panel b: conventional persistence diagram that highlights the single dominant 2-cycle (vertically highest blue square) expected for the dataset. Panel c: flat persistence diagram clearly highlights the single dominant 2-cycle (vertically highest blue square). In contrast to (b), this panel effectively uses the entire plot, thus minimizing whitespace and allowing for easier interpretation of more spread out features.}
    \label{fig:sphere}
\end{figure}

The flat persistence diagram addresses the aforementioned issues by dedicating the vertical axis to plotting feature persistence, not the diameter at feature disappearance.
Although persistence is clearly a function of appearance and disappearance, explicitly plotting persistence on the vertical axis more clearly displays useful information to researchers.
Furthermore, it replaces the previously required step of calculating vertical distance between a feature and the diagonal reference line with the simpler and more intuitive step of noting the vertical coordinate.
As an example, we refer to Figure~\ref{fig:sphere}, which uses three plots to visualize the persistent homology of a hollow sphere.
The single dominant 2-cycle seen in all panels is consistent with the homology of a spherical point cloud.
However, to showcase the benefits of the flat persistence diagram, we will focus on the three prominent 1-cycles (green).
Specifically, these are the three 1-cycles with persistence greater than $0.3$ in Figure~\ref{fig:sphere}c.
A clear distinction between Figure~\ref{fig:sphere}b and \ref{fig:sphere}c is the altered order of the three 1-cycles by magnitude of persistence.
In the conventional persistence diagram (Figure~\ref{fig:sphere}b), it appears that the middle 1-cycle (by feature appearance) persists the least of these three features.
In contrast, the flat persistence diagram in panel (c) shows that the middle 1-cycle persists longer than the one to its right.
This is only one example of how the diagonal reference line in conventional persistence diagrams can alter perception and prevent accurate interpretation of persistent homology.
This inherent visual bias lends support for increased utilization of the flat persistent diagram, which eradicates bias by directly plotting the most relevant property of features (persistence) on the vertical axis.

In addition to decreasing visual bias, the flat persistence diagram uses space more efficiently than the conventional persistence diagram.
In Figure~\ref{fig:sphere}b, all features are squeezed into the top-left half of the plot with a significant amount of empty space on the lower-right half, a direct corollary of the feature disappearance necessarily being strictly greater than feature appearance. 
In contrast, the flat persistence diagram in panel (c) allows for better use of this otherwise unused space.
The plotted features in Figure~\ref{fig:sphere}c are more spread out, and are thus easier to interpret and more aesthetically pleasing, than the features in \ref{fig:sphere}b.
Thus, flat persistence diagrams clearly grant a useful purpose to the untouched whitespace that usually comprises one half of conventional persistence diagrams.

\section{Discussion}
\label{sec:discuss}

The problem of computing substrings provides an example from the field of computer science analogous to the proposed flat persistence diagram.
Consider the character string \texttt{Hello}.
If programmers wish to extract the substring \texttt{ell} they have (at least) two distinct ways of thinking about the process.
One option is to think of the relevant substring as starting at the 2nd character and ending at the 4th.
Alternatively, one can think of the relevant substring as starting at the 2nd character and being 3 characters in length.
Both methods return equal substrings, and are each useful frameworks of thought for different contexts.
Similarly, a feature can be thought of as appearing at $d=2.0$ and disappearing at $d=4.0$.
Alternatively, it can be considered to appear at $d=2.0$ and persist for $2.0$ units.
The distinction between these two approaches lies in the focus on either feature disappearance or feature persistence.
Although feature disappearance could aid in characterization of a dataset's shape, we believe that in the context of topological data analysis, the focus of persistent homology visualization is predominantly geared toward identifying features with significant persistence.
This goal is better achieved through use of flat persistence diagrams than the corresponding conventional plot.

Visualizing a one-dimensional, quantitative variable is a basic statistical task.
However, a variety of plots (e.g. histograms, density plots, and boxplots) are selected from depending on the specific context of this simple task.
Similarly, persistence diagrams (flat and conventional) and persistence barcodes are useful visualizations, however, the lack of development of additional alternate visualizations of persistent homology could be hampering research in the emerging field of topological data analysis by preventing researchers from gleaning maximal knowledge from their work.

Visualization of persistent homology is still in early stages, with the first iteration of two early forms of visualization proposed in the infancy of topological data analysis in widespread use today.
We have identified clear deficiencies in the persistence diagram and proposed an alternative - the flat persistence diagram - that corrects inherent issues without significant concurrent downsides.
As discussed in Section~\ref{sec:review}, there remain inherent limitations in visualizing persistent homology with persistence diagrams and persistence barcodes.
Thus, research into visualizations for more complex forms of persistent homology~\citep{rivet} should be paralleled by continued research into improved visualizations for simple persistent homology.
Further progress in data visualization can reduce the aforementioned limitations and help researchers elucidate more topological insights from datasets with the aid of a greater variety of visualizations.

\section*{Acknowledgments}
\label{sec:ack}

RRW thanks Elena Svenson, PhD for insightful conversation.
The authors also thank the Cleveland Clinic Foundation and Case Western Reserve University for research support, and the members of Theory Division at the Cleveland Clinic's Lerner Research Institute for camaraderie.

\bibliographystyle{plainnat}
\bibliography{references}

\end{document}